# SiC Graphene Suitable For Quantum Hall Resistance Metrology.

Samuel Lara-Avila[1], Alexei Kalaboukhov[1], Sara Paolillo[2], Mikael Syväjärvi[3], Rositza Yakimova[3], Vladimir Fal'ko[4], Alexander Tzalenchuk[5], Sergey Kubatkin[1]

[1]Department of Microtechnology and Nanoscience, Chalmers University of Technology, S-412 96 Göteborg, Sweden

[2]Department of Physics, Politecnico di Milano, 20133 Milano, Italy

[3]Department of Physics, Chemistry and Biology (IFM), Linköping University, S-581 83 Linköping, Sweden

[4]Physics Department, Lancaster University, LA1 4YB, UK

[5]National Physical Laboratory, TW11 0LW Teddington, UK

Discovery of graphene – a single layer of graphite – raised hopes that the roadmap for future will extend beyond the limitations of the silicon era technologies. The distinctiveness of graphene as an electronic material is determined by the linear dispersion of charge carriers, $\pm vp$, which mimic the behaviour of 'relativistic' two-dimensional electrons with the Dirac velocity $v \approx 10^8$ cm/s [1]. The recently observed Klein tunnelling of electrons through the *p-n* junctions is one of several spectacular manifestations of graphene's unique electronic properties [2], while an unusual sequencing of the quantum Hall effect (QHE) plateaux at the values $R_{xy}^{(n)} = \frac{h}{(4n+2)e^2}$ [1] is widely considered to be a true smoking gun for the chiral electrons in this material.

This impressive range of unconventional transport properties of electrons in graphene have, by now, been seen only in the flakes mechanically exfoliated from bulk graphite. Perfectly suitable for attaining a small number of extremely high-quality structures, the exfoliation process is laborious and, thus, difficult to implement in the mass assembly of large-scale integrated electronic circuits. An alternative 'top-down' approach to the use of graphene in electronics consists of growing it epitaxially [3]. To this end, there is an open fundamental question: to what extent the synthesized layer retains the unique electronic properties of graphene. Although angle-resolved photoemission studies of such material [4] have revealed an almost linear dispersion of carriers around the corners of a hexagonal Brillouin zone typical for graphene, up to now no observation of the QHE in epitaxial graphene has been made.

Here we report the observation of the quantization of Hall resistance peculiar to graphene in devices made of graphene synthesized on the Si-terminated surface of SiC. In these measurements we clearly identify two QHE plateaux, at $R_{xy}^{(0)} = \frac{h}{2e^2} \approx 12906.4\,\Omega$ and $R_{xy}^{(1)} = \frac{h}{6e^2}$ corresponding to the filling factors $\nu = 2$ and $\nu = 6$, respectively. We also demonstrate that the quantisation of Hall resistance and vanishing of dissipative conductivity at the $\nu = 2$ plateau has, at least, 5-digit accuracy at cryogenic temperatures, T= 0.3 – 4 K. In graphene $\nu = 2$ corresponds to the fully occupied zero-energy Landau level characterised by the largest separation $\hbar v \sqrt{\hbar c / eB}$ from other Landau levels in the spectrum.

The material studied in these experiments has been grown in 1 atm Ar at about 2000°C on the Si-terminated face of well oriented 4H-SiC substrate [5]. The reaction kinetics on the Si-face is slower than on the C-face because of the higher surface energy, which helps homogeneous and well controlled graphene formation [6]. Graphene was grown at 2000ºC and 1 atm Ar gas pressure, which result in an average terrace width of about 6 μm and step height of 10-15Å. Monolayers of graphene were atomically uniform over more than 50 μm². 20 Hall bar devices of different sizes, from 160 μm x 35 μm down to 5.8 μm x 1 μm, were produced on each wafer using standard electron beam lithography and oxygen plasma etching (Figure 1A&B). From low-field measurements, the manufactured material is *n*-doped, due to the charge transfer from SiC [4,6], with the measured electron concentration in the range of (5-8)x10$^{11}$ cm$^{-2}$, mobility about 2400 cm²/Vs at room temperature and between 4000 and 7500 cm²/Vs at 4.2 K almost independent of device dimensions.

Figure 1C shows typical behavior for the manufactured devices of the longitudinal (dissipative) $R_{xx}$ and the transverse (Hall) $R_{xy}$ resistance of a 2 μm wide Hall bar at 4.2 K and magnetic field up to 14 T. At high magnetic field $R_{xy}$ forms a plateau at $h/2e^2$, with the instrumentation-limited uncertainty better than 100 ppm, accompanied by a vanishing $R_{xx}$. The plateau onset appears in the field range field 9-12 T, depending on the carrier concentration, which is beyond our control in this experiment. Stronger doping leads to a higher onset field. The $h/6e^2$ plateau is not so flat, and $R_{xx}$ develops only a weak minimum. However, its presence confirms that the studied material is indeed monolayer graphene. In low magnetic fields we observe Shubnikov-de Haas oscillations as well as characteristic features of phase coherence of electrons in a disordered conductor: a weak localization peak and reproducible conductance fluctuations within the interval ±2 T.

The reported observation of the QHE in graphene synthesized on SiC fills a yawning gap in the understanding of the electronic properties of this material. Despite numerous efforts, QHE was not observed in earlier studies of epitaxial graphene on SiC. The difficulty was related to the lack of atomically accurate thickness control during film growth (especially on the carbon-terminated facet) and a strong variation of carrier density (doping) from the bottom to the top graphitic layers. Having found the QHE in several devices produced on distant parts of a single large-area wafer, we can confirm that material synthesized on the Si-terminated face of SiC promises a suitable platform for the implementations of quantum resistance metrology at elevated temperatures and, in the longer term, opens bright prospects for scalable electronics based on graphene.

Figure captions:

Figure 1.   A. Layout of a 7x7 mm wafer with 20 patterned devices. One of the devices wire-bonded to the chip carrier. Pairs of current and voltage electrodes are marked.
B. Close-up of the wire-bonded device with L=11.6 µm and W=2 µm.
C. Transverse ($R_{xy}$) and longitudinal ($R_{xx}$) resistance measured on this device at T=4.2 K across the pairs of contacts ($V_1+V_1-$) and ($V_2+ V_1+$) respectively with 1 µA current between I+ and I-.

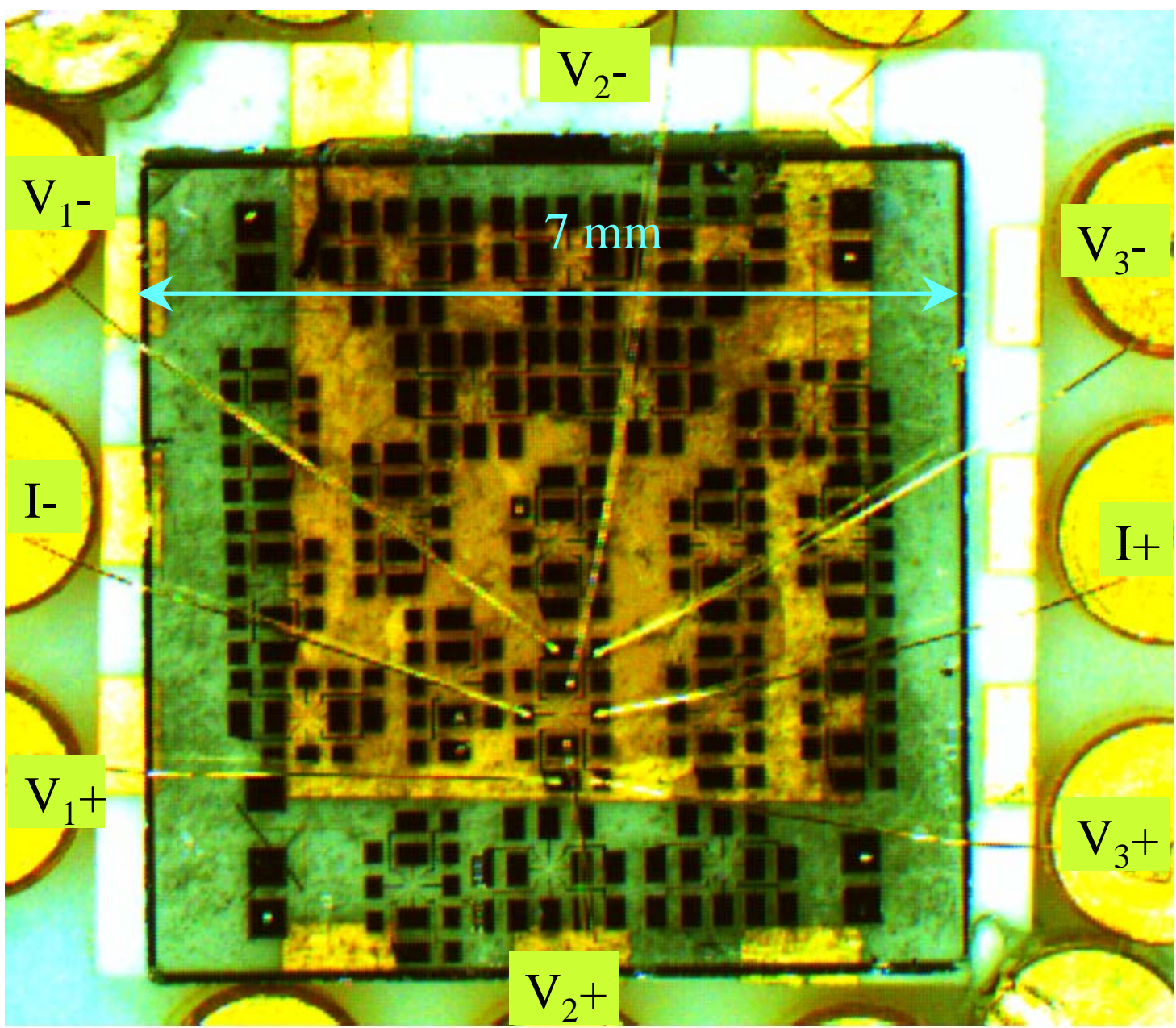

Figure 1A

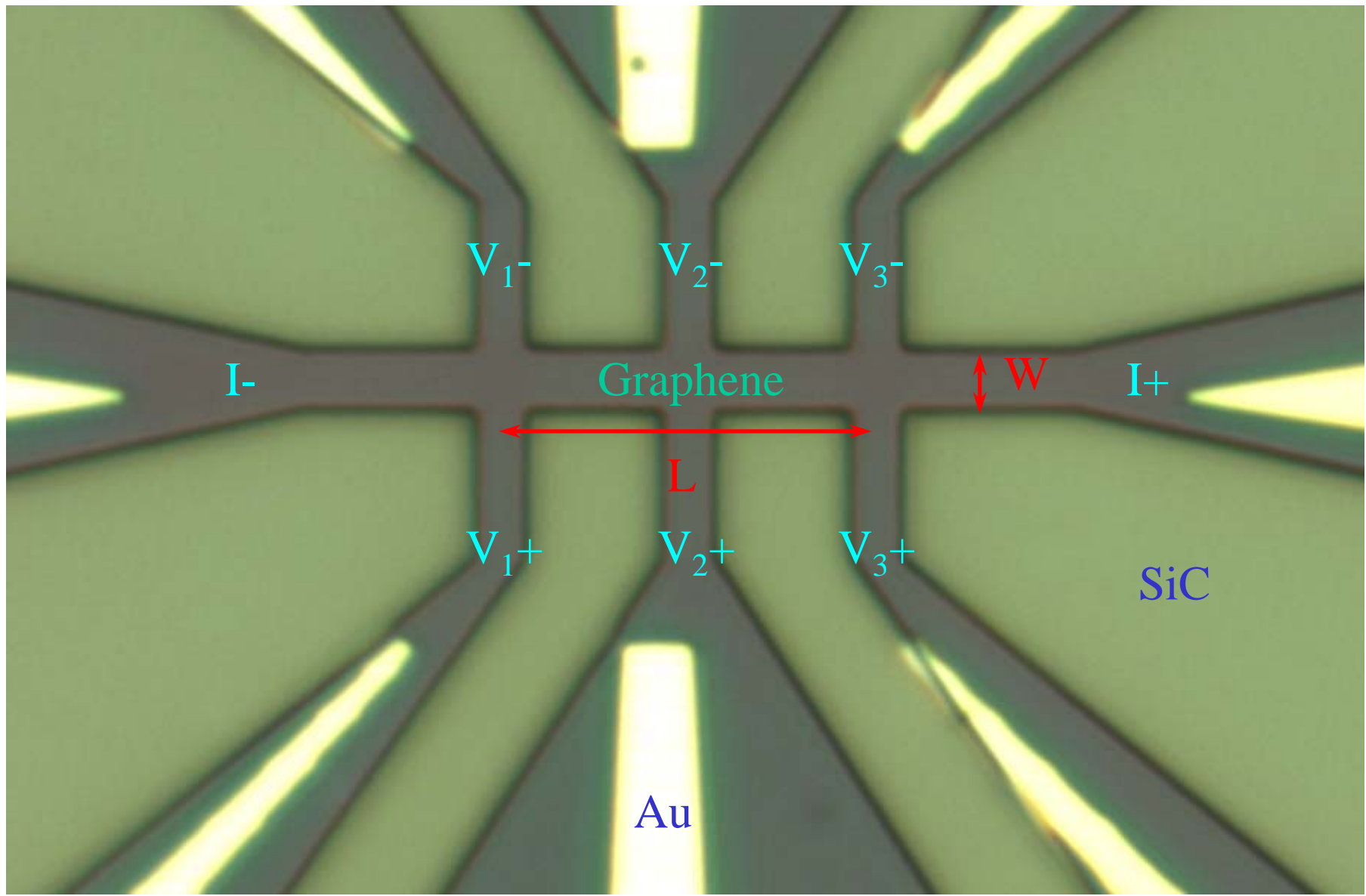

Figure 1B

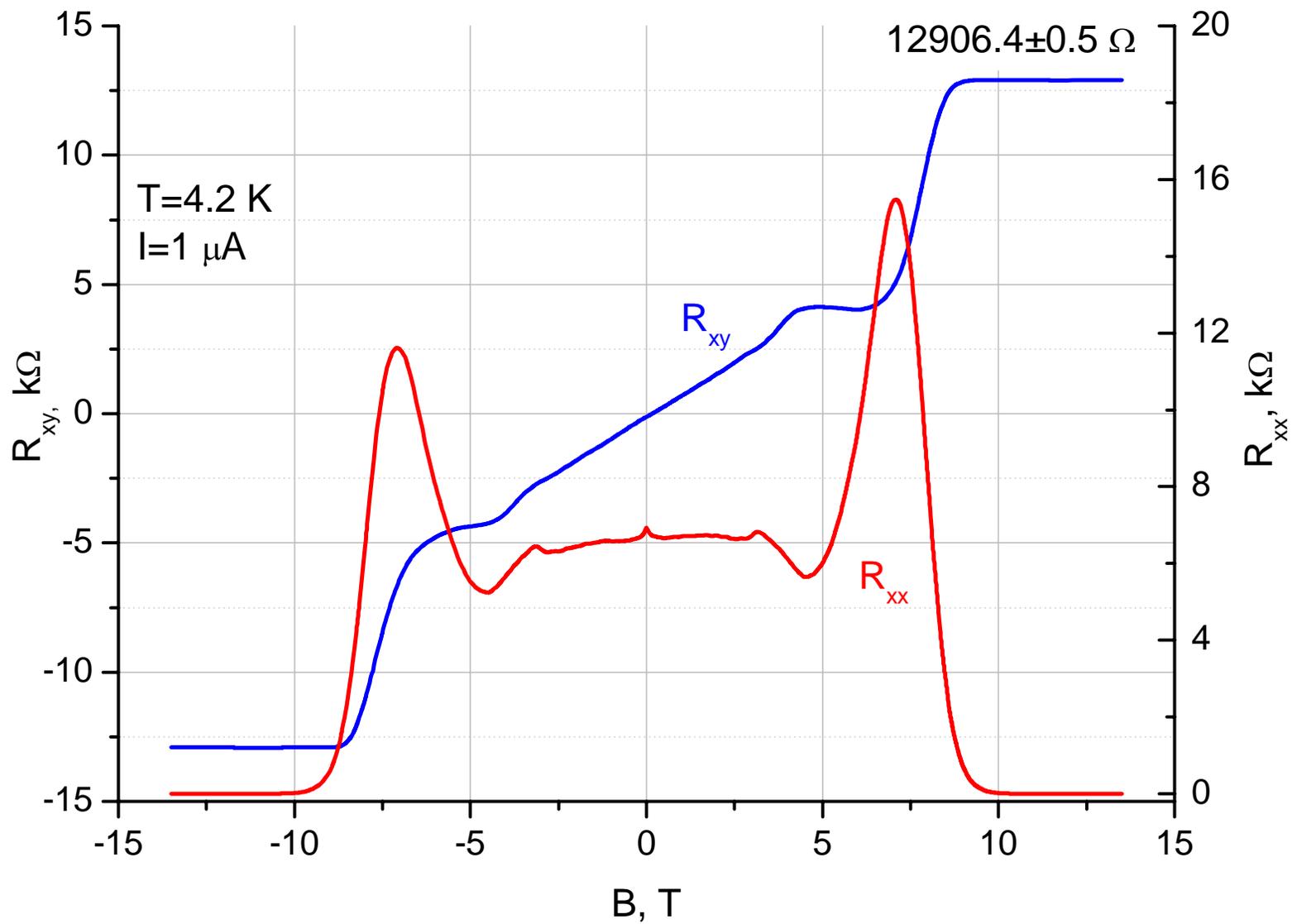

Figure 1C